\documentclass[aps,amsfonts,nofootinbib]{revtex4}
\usepackage{epsfig}
\usepackage{graphicx}
\usepackage{amsmath}
\usepackage{amsbsy}
\usepackage{color}
\usepackage{bbm}

\newcommand{\ee}{\end{equation}}
\newcommand{\bb}{\begin{equation}}
\newcommand{\eqb}{\begin{eqnarray}}
\newcommand{\eqf}{\end{eqnarray}}

\begin{document}
\title{Non-Abelian   Monopoles  as   the   Origin   of  Dark   Matter}
\author{H.    Falomir}
\email{falomir@fisica.unlp.edu.ar}
\affiliation{IFLP    -   CONICET,
  Departamento de  F\'{\i}sica, Universidad  Nacional de La  Plata, La
  Plata, Argentina}
\author{J.  Gamboa}
\email{jgamboa55@gmail.com}
\affiliation{Departamento  de  F\1sica,  Universidad  de  Santiago  de
  Chile, Casilla 307, Santiago, Chile}
\author{F.   M\'endez}
\email{fernando.mendez@usach.cl}
\affiliation{Departamento de F\1sica,
Universidad de Santiago de Chile, Casilla 307, Santiago, Chile}

\begin{abstract}
  We suggest that dark matter may be partially constituted by a dilute
  't Hooft-Polyakov monopoles gas. We reach this conclusion by using the
  Georgi-Glashow model  coupled to a dual kinetic mixing term $ F{\tilde {\cal G}}$ where $F$ is the electromagnetic field and ${\cal G}$ the 't Hooft tensor.  We show that these monopoles carry both (Maxwell) electric and (Georgi-Glashow) magnetic charges and the electric charge quantization condition is modified in terms of a dimensionless real parameter. This parameter could be determined from milli-charged particle experiments.
    \end{abstract}

\maketitle


\eject


The detection of  dark matter is one of the  most important challenges
on high  energy physics  in present days  because its  discovery would
explain a number of very  important unsolved problems in astrophysics,
astronomy and  particle physics \cite{review1}. Since  the dark matter
interacts  very weakly  with visible  matter, it  seems that  the most
promissory  way  to  detect  it  would  be  through  indirect  methods
as, for example, the  detection of the products of the
annihilation of pairs of dark matter/anti-dark matter
which  could  produce  overabundance   of  visible  matter  or  energy
\cite{indirect}.
Direct detection  methods, on the  other hand, has also  been proposed
and it is  a very active field of research \cite{direct}.
\\

The expected overabundance of visible matter could be explained
via annihilation of dark matter ($\chi$) processes of the type
\[
\chi+{\bar \chi} \to \text{visible matter}
\]
whose  evaluation  requires,  of  course,  the  precise  knowledge  of the
production mechanism, so far unknown. The  excess of gamma rays at the
galactic  center is,  for example,  a possible  signal of a dark matter
annihilation process \cite{phexcess}, and other effects \cite{review2} might be understood as
a manifestation of some relevant nonperturbative mechanism presumably not
taken into account until now.

\vskip 0.15cm

In this letter  we would like to explore a  model in which the
dark  sector contains  a  non-abelian field  theory admitting  massive
topologically stable  classical solutions: monopoles  which weakly
interact with ordinary fields.  These configurations are characterized by a \emph{topological charge}, an additive magnitude which can take both signs in such a way that two such solutions can be smoothly merged into a single object with the sum of their topological charges. We will assume  that it is energetically favorable to have configuration with the same topological charge sign separated far away. Then, we can consider a dilute gas of monopoles of unit topological charge (of both signs), neutral in the mean, created at a very energetic event in the past.

\vskip 0.15cm

Under these conditions, the most relevant interaction for such objects
would  be  the  annihilation  of   a  monopole  (charge  +1)  with  an
anti-monopole  (charge -1),  producing  a  non-topological object  and
emitting  their energy  in the  form of  dark and  ordinary particles.
Being  very  massive,  these  objects could also be  gravitationally
attracted by the galaxies, enhancing
%
the probability of such annihilation process in their immediacies.  If
the result includes the emission  of normal photons, this scheme would
fit in the excess of luminosity of galaxies centers \cite{phexcess}.
\vskip 0.15cm

There is also a simple  experimental example which provides an analogy
of the aforementioned scenario. Indeed, in
a ripple  tank with two vortices  produced in the water  with opposite
sense of  rotation, the collision between  these two \emph{topological
  defects} produces a disturbance in water  in the form of waves whose
amplitude will depend on the vortex energies
\cite{berry,melo}, in virtue of the
energy and angular momentum conservation.

\vskip 0.15cm

Our discussion is  based on  a simple
application of the Georgi-Glashow model  \cite{GG} for the dark sector
with a suitable gauge-invariant \emph{kinetic mixing term} between the
nonabelian  (dark) fields  and  the usual  electromagnetic field.  The
coupling between both sectors is  realized by adding to the Lagrangian
a term proportional  to the electromagnetic strength  tensor times the
gauge-invariant 't Hooft tensor \cite{thooft,polyakov,weinberg}.


In order  to establish our  notation, let  us start by  describing the
Lagrangian of  the Georgi-Glashow  model, an $SU(2)$  nonabelian gauge
theory with a triplet of scalar fields in the adjoint representation,
\bb
{\cal   L}_{GG}   =   -\frac{1}{4}   G_{a,\mu   \nu}(X)G_a^{\mu   \nu}
(X)-\frac{1}{2}  (D_\mu[X]   \phi)_a  (D^\mu  [X]\phi)_a   -  V(\phi_a
\phi_a),
\label{1}
\ee
where an implicit summation over  $a, b, ...=1,2,3$ is understood each
time an index is repeated in a term. Here, the covariant derivative of
the scalars in the adjoint representation is given by
\eqb
(D_\mu  \phi)_a &=&  \partial_\mu \phi_a  + q\,   \epsilon_{abc}X_{b,\nu}
\phi_c.
 \label{3}
\eqf
where $X_{a,\mu}$  is the  (nonabelian) gauge  field and  the strength
tensor is given by
\eqb
G_{a,\mu \nu}(X)&=& \partial_\mu X_{a,\nu} -\partial_\nu X_{a,\mu} + q\,
\epsilon_{abc} X_{b,\mu} X_{c,\nu},
\label{2}
\eqf
where    $\epsilon_{abc}$    is    completely    antisymmetric    with
$\epsilon_{123}=1$.

Notice that both the strength tensor and the triplet of scalars can be
represented as elements in the Lie algebra of $SU(2)$ as
\bb
G_{\mu \nu} = G_{a,\mu \nu} T_a\,, \quad \Phi= \phi_a T_a\,,
\ee
where $T_a=  \frac{\sigma_a}{2}, a=1,2,3$, are the  $SU(2)$ generators
which satisfy  $\left[ T_a,T_b  \right]= i \epsilon_{a  b c}  T_c$ and
$\rm{tr} \left\{T_a T_b\right\}=\frac{\delta_{a b}}{2}$. Under a gauge
transformation $U(x)\in SU(2)$ these elements transform as
\bb
G_{\mu  \nu}  \rightarrow  U   G_{\mu  \nu}  U^\dagger\,,  \quad  \Phi
\rightarrow U \Phi U^\dagger.
\ee

We will  assume that  the potential $V(\phi_a  \phi_a)\geq 0$  has its
absolute  minima  at  $\phi_a   \phi_a=v^2$.  Then,  the  symmetry  is
spontaneously broken  to $U(1)$.   We also  adopt the  temporal gauge,
$X_{a,0}=0$.

We  will   be  interested   in  \emph{background}   nontrivial  static
configurations  of finite  energy, condition  which requires  that the
scalar fields tend to a minimum of the potential sufficiently fast for
$\textbf{x}^2=r^2  \rightarrow  \infty$.  Since the  potential  minima
belong to  a 2-sphere (of  radius $v$), $\mathcal{S}_v^2$,  the scalar
field     at     infinity     establishes    an     application     of
$\mathcal{S}_\infty^2= \partial  \mathbb{R}^3$ onto $\mathcal{S}_v^2$,
which  is characterized  by the  \emph{winding number},  the (integer)
number  of  times  the  application  involves  the  sphere  of  minima
$\mathcal{S}_v^2$,  with  a  sign  determined by  the  sense  of  this
covering. 't\  Hooft \cite{thooft}  and Polyakov  \cite{polyakov} have
shown that  there exist such  static and finite energy  solutions with
nontrivial winding number, which are  {\it stable} as a consequence of
their topology.

't\ Hooft \cite{thooft} has also  constructed a gauge invariant tensor
given by
\bb
\begin{array}{c} \displaystyle
  {\cal G}_{\mu \nu}  =2 \, {\rm tr} \left\{ G_{\mu  \nu} \hat{\Phi} +
  \frac{i}{q} \left[ D_\mu \hat{\Phi},  D_\nu \hat{\Phi} \right] \hat{
  \Phi} \right\} = \\ \\
\displaystyle
  = G_{a,\mu \nu}\,  {\hat \phi}_a - \frac{1}{q}  \, \epsilon_{abc} \,
  {\hat \phi}_a \, (D_\mu {\hat \phi})_b\, (D_\nu {\hat \phi})_c ,
\label{abel}
\end{array}
\ee
where    $\hat{   \Phi}=\hat{\phi}_a    T_a$,   with    $\hat{\phi}_a=
\phi_a/\sqrt{\phi_b \phi_b}$.  Notice that  $ {\cal G}_{\mu  \nu}$ has
dimension of  $(mass)^2$.  This tensor  can be brought to  coincide with
$G_{3,\mu\nu}$,  for   example,  in   any  \emph{bounded}   region  of
$\mathbb{R}^3$  through  a  suitable  (smooth)  gauge  transformation,
without changing the winding  number. These considerations justify the
interpretation   of    $\mathcal{B}_i:=\frac{1}{2}\epsilon_{i   j   k}
\mathcal{G}_{j k}$  as the  \emph{magnetic field} associated  with the
unbroken $U(1)$ symmetry \cite{weinberg}, and
\bb
g:= \frac{1}{8\pi} \oint_{\mathcal{S}_\infty^2} \mathcal{B}_k\,   dS_k =
 \frac{-1}{8\pi    q}   \,    \epsilon_{i   j    k}\,   \epsilon_{abc}
 \oint_{\mathcal{S}_\infty^2}   {\hat   \phi}_a  \,   \partial_\mu   {
   \phi}_b\, \partial_\nu {\hat \phi}_c \, dS_k
\ee
as the \emph{magnetic charge} of the topological configuration of this
nonabelian field.

So      defined,     $g$      is      a     topological      invariant
\cite{thooft,polyakov,weinberg}  which   equals  the   winding  number
divided by  the constant $q$ \cite{arafune},  $g=n/q$. Moreover, since
it  is a  surface  integral,  $g$ is  an  additive  quantity for  well
separated topological configurations, and  its value remains invariant
when  these  configurations are  smoothly  brought  together. In  this
sense,  these classical  configurations of  the nonabelian  theory are
\emph{monopoles}  of  \emph{magnetic  charge} quantized  in  units  of
$q^{-1}$.

The static solution with $n=\pm 1$ is the 't\, Hooft-Polyakov monopole
\cite{thooft,polyakov},   a   regular   configuration  free   of   the
\emph{Dirac singularities} present in  the description of monopoles in
an Abelian gauge theory. These are \emph{massive} configurations which
minimize the energy functional given by
\bb
E[\mathbf{X},\Phi]=\int_{\mathbb{R}^3}      \left\{      \frac{1}{2}\,
  G_{a,ij}G_a^{ij} +\frac{1}{2}\,  {(D_i \phi)_a}^2  + V(\phi_a\phi_a)
\right\} d^3x\,,
\ee
with $V(\phi_a\phi_a)=\frac{\lambda}{8}\left(\phi_a\phi_a-v^2\right)^2$.

The  numerical  solution  of   the  variational  problem  for these minima \cite{preskill} shows  that
 there   is  a   \emph{core}  of   radius  $R_c\approx
{M_X}^{-1}=1/q  v$  outside which  the  massive  gauge bosons {(of mass $M_X=q v$)}  rapidly
approach their asymptotic value, while the scalar field approaches its
asymptotic     value     outside     a     region     of     dimension
$R_H={M_H}^{-1}=1/v\sqrt{\lambda}$ {(where ${M_H}=v \sqrt{\lambda}$ is the scalar (\emph{Higgs}) mass)}, less  than $R_c$  for sufficiently
large $\lambda$.  The main contributions  to the energy come  from the
\emph{magnetic field} $\mathbf{\mathcal{B}}$ outside the core and from
the  gradient of  the scalar  field inside  it. 't\,  Hooft has  shown
\cite{thooft,preskill} that the monopole mass is
\bb
 \label{mass}
M_{\mbox{\tiny{mon}}}= \frac{4\pi v}{q}\,  f(\lambda/q^2)= \frac{4\pi}{q^2 } \,
M_X f(\lambda/q^2)\,,
\ee
where   $f(x)$  is   an  $O(1)$   monotonically  increasing   function
\cite{preskill}, $1\leq f(x) <2$ for $x\in \mathbb{R}^+$.

\vskip 0.25cm

After this  brief description of known  results about one of  the most
relevant non-perturbative  developments obtained in the  seventies, we
proceed to employ  this nonabelian gauge theory as a  model for a dark
matter sector,  which we  put in  interaction with  the \emph{visible}
Maxwell electromagnetic  field $A_\mu$ by  adding to the  Lagrangian a
(gauge invariant) {\it dual kinetic mixing term} \cite{holdom}
 \bb
{\cal L}_1  = \gamma \,  F_{\mu \nu}  (A) {\cal {\tilde  G}}^{\mu \nu}
(X).
\label{new}
\ee
Here $\gamma$ is a dimensionless constant, $F_{\mu \nu}  (A)$ is the strength tensor
of the electromagnetic field and the dual strength tensor of the nonabelian field
is ${\cal  {\tilde G}}^{\mu \nu} (X)  = \frac{1}{2}\,\epsilon^{\mu \nu
  \rho \sigma} {\cal G}_{\rho \sigma}  (X)$, with ${\cal {G}}_{\mu \nu}
(X)$ defined in (\ref{abel}).

It would be noticed that this coupling between the dark sector and the electromagnetic field breaks the CP-symmetry but, as we will see, it also gives rise to the possibility of having massive milli-charged objects whose detection would be an indication in favor of this model. In fact, an independent argument  of indirect  CP-violation in the dark sector has been recently proposed in \cite{paper1}.

Our Lagrangian is then
\eqb
{\cal L}_0  &=& {\cal L}_{Maxwell}  + {\cal L}_{GG}+ \gamma  \, F_{\mu
  \nu} (A) {\cal {\tilde G}}^{\mu \nu} (X)
\nonumber
\\
&=&  -\frac{1}{4}  F_{\mu  \nu  } (A)  -\frac{1}{4}  G_{a,\mu  \nu}(X)
G_a^{\mu \nu} (X) - \frac{1}{2}  (D_\mu[X] \phi)_a (D^\mu [X]\phi)_a -
V(\phi_a \phi_a)  + \gamma \,  F_{\mu \nu} (A) {\cal  {\tilde G}}^{\mu
  \nu} (X),
\label{mod12}
\eqf
where we  will assume  that the  visible and  dark sectors  are weakly
coupled ($\gamma \ll 1$).

It  is worth  to remark  that  in \cite{BJ}  a similar  model of  dark
monopoles   in  interaction   with   the  visible   sector  has   been
considered. In  that article,  the interaction  does not  break parity
and,  as  a consequence,  the  monopoles  are  not  a source  for  the
electromagnetic field. In the following we will show that the coupling
we propose  in \eqref{new}  makes the monopoles  to appear  as massive
charged  objects,  which  effectively  are sources  for  the  visible
electromagnetic field \cite{others1}.

Indeed,  the Euler-Lagrange  equations for  the electromagnetic  field
derived from Ec.\ \eqref{mod12} are
\bb
\partial_\mu        \frac{\partial       L}{\partial\left(\partial_\mu
    A_\nu\right)}  =  \partial_\mu\left\{  -  F^{\mu\nu}  +  2  \gamma
  \tilde{\mathcal{G}}^{\mu\nu} \right\}=0\,,
\ee
which implies that  the content of electric charge  of the topological
configuration in Eq.\ \eqref{abel} is
\bb
Q_{\mbox{\tiny{mon}}}= \int_{\mathbb{R}^3} \partial_i F^{i0} \, d^3 x = \gamma
\epsilon^{i0jk} \int_{\mathbb{R}^3} \partial_i \mathcal{G}_{jk} \, d^3
x
=  2 \gamma  \oint_{\partial\mathbb{R}^3}  \mathcal{B}_i \,
dS_i = 16\pi\gamma g= \frac{16\pi\gamma}{q} \, n\,.
\label{charge}
\ee
Therefore, these monopoles also  carry an electric charge proportional to
the winding number, quantized in units of ${16\pi\gamma}/{q}$.

\medskip

The next  step is to physically  interpret the model, in  which we are
essentially assuming that the  electromagnetic field is weakly coupled
to the  nonabelian sector ($\gamma <<  1$) and the monopoles  are very
massive background  configurations, which  requires that $q  <1$ (See
Eq.\ \eqref{mass}). On the other hand, the monopole electric charge in
Eq.\ \eqref{charge} must be small  in order they remain \emph{dark} to
the  electromagnetic interaction;  then,  $0<\gamma <<  q<1$. Taking  into
account the  additivity of  the magnetic charge,  we have  also assumed
that it is energetically favorable to have monopoles of winding number
$\pm 1$.

These  monopoles  are  topologically stable  classical  configurations
which cannot individually decay  through the emission of dark  or visible particles,
since these processes do not  change their winding number.  Its decay
can only  occurs when a pair  monopole - antimonopole meet  each other
and, due to the additivity of  the magnetic charge, they merge into an
electrically neutral  configuration with vanishing winding  number. In
this case, an  energy equal to twice the monopole  mass can be emitted
as visible and dark particles. This possibility presents as an  interesting route to explore in the
context of  the observed  photon excess at the center of  the galaxy
\cite{phexcess}.

The  precise mechanism  of annihilation  of dark  matter still  remains
unknown, but  the possibility  that  it decays  in  cascades
 until  finally reaching a pair of particle-antiparticle
of the standard model is an interesting prospect
to explore in order to get some numerical bounds.
In this context, it would be worth to consider the model previously discussed.

{{As previously mentioned, the CP-breaking kinetic mixing approach in Eq.\ \eqref{new} for the interaction between the Maxwell field and a nonabelian $SU(2)$ gauge theory for the dark sector leads to the appearance of 't Hooft - Polyakov monopoles in the dark sector which, additionally to their (nonabelian) mgnetic charge, present in the visible sector as millicharged massive particles, whose detection would give support for this proposal}} \cite{cline}.

  \vskip 0.25cm
\acknowledgements
We would like to thank Prof. J. L. Cort\'es  and M. Tytgat by discussions.
This work was supported by FONDECYT/Chile grants 1130020 (J.G.), 1140243 (F.M.).
H.F.  thanks ANPCyT, CONICET  and UNLP, Argentina, for partial support through grants
{PICT-2011-0605}, {PIP-112-2011-01-681} and {Proy.\ Nro.\ 11/X615} respectively.


\begin{thebibliography}{99}
\bibitem{review1}  Particle  Dark  Matter:  Observations,  Models  and
  Searches, G.  Bertone (Ed.), Cambridge University  Press (2013). See
  also {\it e.g.} G.~Bertone, D.~Hooper and J.~Silk,
  Phys.\ Rept.\ {\bf 405}, 279 (2005).


\bibitem{indirect} See {\it e.g.}, F.  Donato, Phys.  Dark Univ.  4, 41 (2014) (DARK
  TAUP2013); T. Bringmann, Pos EPS-HEP2011, 061 (2011).

\bibitem{direct} See the review on the subject by T. M. Undagoitia and
  L. Rauch, J. Phys.  G {\bf 43}, 1 (2016) and references therein.




\bibitem{phexcess} L. Goodenough and D. Hooper,
arXiv:0910.2998; D. Hooper and L. Goodenough,
Phys.  Lett. B  {\bf 697}  412  (2011). For  a recent  review see  for
example  F. Calore,  I.   Cholis,  C.  McCabe  and  C. Weniger,  Phys.
Rev.  D  {\bf 91},  063003  (2015);  F.  Calore,  I.  Cholisb  and  C.
Weniger, JCAP 1503, 038 (2015).

%
%

\bibitem{review2} See {\it e.g.}  D.~S.~Akerib {\it et al.} [LUX Collaboration],
  Phys.\ Rev.\ Lett.\  {\bf 112}, 091303 (2014)
  doi:10.1103/PhysRevLett.112.091303
  [arXiv:1310.8214 [astro-ph.CO]].


  \bibitem{berry} M. V. Berry, R. G. Chambers, M. D. Large, C. Upstill
    and J. C. Walmsley, Eur. \ J. \ Phys. {\bf 1}, 154 (1980).


  \bibitem{melo} F.  Vivanco, F. Melo,  C. Coste, and F.  Lund, Phys.\
    Rev. \ Lett. {\bf 83}, 1966 (1999).


\bibitem{GG} H.~Georgi and S.~L.~Glashow,
  Phys.\ Rev.\ Lett.\  {\bf 28}, 1494 (1972).



\bibitem{thooft} G.\ 't Hooft, Nucl.\ Phys.\ \textbf{B79}, 276 (1974).


\bibitem{polyakov}A.M.\ Polyakov, JETP Letters \textbf{20}, 194 (1974).




\bibitem{weinberg} S.\  Weinberg, \emph{The quantum Theory  of Fields,
    Vol.\ II:  Modern Applications}, Cambridge University  Press, USA,
  1996.


\bibitem{arafune}J.\  Arafune, P.G.O.\  Freund and  C.J.\ Goebel,  J.\
  Math.\ Phys.\ \emph{16}, 433 (1975).


\bibitem{preskill} See {\it e.g.}  J.~Preskill,
  Ann.\ Rev.\ Nucl.\ Part.\ Sci.\  {\bf 34}, 461 (1984).



\bibitem{holdom} B. Holdom, Phys. Lett. B 166, 196 (1986). See also, M. Pospelov, Phys. Rev. D {\bf 80}, 095002 (2009); M. Pospelov and A. Ritz, Phys. Lett. B
{\bf 671}, 391 (2009); M. Pospelov, A. Ritz and M. B. Voloshin, Phys. Lett. B {\bf 662}, 53 (2008).








\bibitem{paper1} W.~Chao, M.~J.~Ramsey-Musolf and J.~H.~Yu,
  arXiv:1602.05192 [hep-ph].


\bibitem{BJ} F.\ Br\"{u}mmer and J.\ Jaeckel,
  Phys.\ Lett.\ B {\bf 675}, 360 (2009); F.~B\"rummer, J.~Jaeckel and V.~V.~Khoze,
  JHEP {\bf 0906}, 037 (2009).

  \bibitem{others1}  See {\it e.g}  V.~V.~Khoze and G.~Ro,
  JHEP {\bf 1410}, 61 (2014),
  [arXiv:1406.2291 [hep-ph]];  C.~G.~Sanchez and B.~Holdom,
  Phys.\ Rev.\ D {\bf 83}, 123524 (2011),  [arXiv:1103.1632 [hep-ph]].

\bibitem{cline} See {\it e.g.} J.~M.~Cline, Z.~Liu and W.~Xue,
  Phys.\ Rev.\ D {\bf 85}, 101302 (2012),
  [arXiv:1201.4858 [hep-ph]];  J.~Jaeckel, J.~Redondo and A.~Ringwald,
  Phys.\ Rev.\ Lett.\  {\bf 101}, 131801 (2008), [arXiv:0804.4157 [astro-ph]]; H.~Gies, J.~Jaeckel and A.~Ringwald,
  Europhys.\ Lett.\  {\bf 76}, 794 (2006),
  [hep-ph/0608238].
  \end{thebibliography}
\end{document}